\documentclass[preprint,eqsecnum,aps]{revtex4} 
\usepackage{graphicx}
\usepackage{dcolumn}

\newcommand{\AmS}{{\protect\the\textfont2
  A\kern-.1667em\lower.5ex\hbox{M}\kern-.125emS}}
\newcommand{\cerenk}{\mbox{\v{C}erenkov}}

\newcommand{\C}{\mbox{$^{12}$C}}
\newcommand{\N}{\mbox{$^{12}$N}}
\newcommand{\numu}{\mbox{$\nu_{\mu}$}}
\newcommand{\numub}{\mbox{$\bar{\nu}_{\mu}$}}
\newcommand{\nue}{\mbox{$\nu_{e}$}}
\newcommand{\nueb}{\mbox{$\bar{\nu}_{e}$}}

\newcommand{\nubx}{\mbox{$\bar{\nu}_{x}$}}
\newcommand{\ep}{\mbox{e$^{+}$}}
\newcommand{\el}{\mbox{e$^{-}$}}
\newcommand{\pos}{\mbox{e$^{+}$}}

\newcommand{\mup}{\mbox{$\mu^{+}$}}
\newcommand{\pip}{\mbox{$\pi^{+}$}}
\newcommand{\mupdecay}{\mbox{\mup\ $\rightarrow\:$ \pos $\!$ + \nue\ + \numub}}

\newcommand{\pipmup}{\mbox{\pip $\rightarrow\:$ \mup + \numu}}
\newcommand{\nuep}{\mbox{\nueb\ + p $\rightarrow$ n + $e^+$ }}
\newcommand{\CC}{\mbox{\C\,(\,\nue\,,\,\el\,)\,\N }}
\newcommand{\CCprot}{\mbox{p\,(\,\nueb\,,\,\ep\,)\,n }}

\newcommand{\nuebx}{\mbox{\nueb $\rightarrow\,$\nubx }}
\newcommand{\numunue}{\mbox{\numu $\rightarrow\,$\nue }}
\newcommand{\numubnueb}{\mbox{\numub $\rightarrow\,$\nueb }}
\newcommand{\NCL}{\mbox{90\%\,C.L.}}
\newcommand{\NnCL}{\mbox{95\%\,C.L.}}
\newcommand{\Dm}{\mbox{$\Delta m^2$}}
\newcommand{\sit}{\mbox{$\sin ^2(2\Theta )$}}
\newcommand{\eVc}{\mbox{eV$^2$/c$^4$}}
\newcommand{\lnL}{\mbox{$\ln \mathcal{L}$}}
\newcommand{\DlnL}{\mbox{$\Delta \ln \mathcal{L}$}}
\newcommand{\Gdng}{\mbox{Gd\,(\,n,$\gamma$\,)}}
\newcommand{\pn}{\mbox{p\,(\,n,$\gamma$\,)}}
\newcommand{\pnd}{\mbox{p\,(\,n,$\gamma$\,)\,d}}

\begin{document}

\preprint{UCRHEP-E330}

\title{Statistical Analysis of Different \numubnueb\ Searches}

\author{
E.D. Church,$^1$\footnote{Now at Prediction Company, Santa Fe, NM, USA.} 
K. Eitel,$^2$\footnote{Corresponding author.}
G.B. Mills,$^3$ 
and M. Steidl $^2$
}
\affiliation{$^1$ University of California, Riverside, CA 92521, USA}
\affiliation{$^2$ Forschungszentrum Karlsruhe, Institut f\"ur Kernphysik, 
             76021 Karlsruhe, Germany}
\affiliation{$^3$ Los Alamos National Laboratory, Los Alamos, NM 87545, USA}

\date{\today}

\begin{abstract}
A combined statistical analysis of the experimental results of the
LSND and KARMEN \numubnueb\ oscillation search is presented. LSND
has evidence for neutrino oscillations that is not confirmed by
the KARMEN experiment. This joint analysis is based on the final 
likelihood results for both data sets. A frequentist approach is 
applied to deduce confidence regions. At a combined confidence level of 36\,\%,
there is no area of oscillation parameters compatible with both 
experiments. For the complementary confidence of $1-0.36=64$\,\%, 
there are two well defined regions of oscillation parameters (\sit ,\Dm) 
compatible with both experiments.
\end{abstract}


\maketitle
%
\section{\label{sec:intro}Introduction}
%
Currently, there are three neutrino anomalies interpreted as evidence for neutrino 
oscillations, namely the atmospheric \cite{atmo}, the solar \cite{solar} and the
LSND \cite{LSNDfinal} anomaly with three distinct squared neutrino mass differences.
However, oscillations between the three Standard Model (SM) neutrinos are described by 
only two independent neutrino mass--squared differences, allowing only two of the above
anomalies as being due to oscillations in a SM scenario extended by massive neutrinos.
Before enlarging the neutrino sector by some additional, unknown {\it sterile} 
neutrinos, each of the experimental results has to be checked independently.

Over the last years, the controversial results of the two
experiments LSND (Liquid Scintillator Neutrino Detector at LANSCE,
Los Alamos, USA) and KARMEN (KArlsruhe Rutherford Medium Energy
Neutrino experiment at ISIS, Rutherford, UK) both searching for
neutrino oscillations \numubnueb\ in a {\it short baseline} regime, 
have led to intense discussions.
The two experiments are similar as they use \numub\ beams from the
\pip -\mup\ decay at rest (DAR) chain \pipmup\ followed by
\mupdecay\ with energies up to 52\,MeV. Furthermore, both
experiments are looking for \nueb\ from \numubnueb\ oscillations
according to the two--flavor oscillation probability \footnote{In a comparison 
of different oscillation searches, one generally has to use a complete 3-- or
4--dimensional description of neutrino masses and mixings, leading to the
general oscillation probability. Since both experiments under investigation here
have similar parameters and, most important, search for oscillations in the
same appearance mode, the simplified formula holds for the purpose
of the following analysis.}
\begin{equation}
 P(\numubnueb) = \sit \sin^2\left(1.27\frac{\Dm\cdot L}{E_{\nu}}\right)
\label{eq:Osci}
\end{equation}
with the difference of the squared masses \Dm\ in \eVc , the flight length
of the neutrino $L$ in meters and the neutrino energy $E_{\nu}$ in units of MeV. 
The detection reaction \CCprot\ provides a spatially correlated delayed
coincidence signature of a prompt \pos\ and a subsequent neutron
capture signal.

This paper describes a combined statistical analysis of the final LSND and
KARMEN\,2 results. There are significant changes to the treatment of LSND data
in the final LSND paper \cite{LSNDfinal}. The present paper uses only the
decay-at-rest (DAR) \numubnueb\ data, and leaves out the impact of the
decay-in-flight (DIF) \numunue\ data. As shown later, the LSND data has 
a favored region in the 7\,\eVc\ region of \Dm\ as a consequence of this change.
This is in contrast to the LSND result \cite{LSNDfinal} which disfavors this solution
at \NCL\ because of the inclusion of the DIF flux.  A motivation for this
change is that a direct comparison of the two experiments is less model
dependent if one considers the LSND DAR flux only, since KARMEN does not 
have a DIF neutrino flux present in its source.

The analysis is based on a frequentist approach
following the suggestions of \cite{feldcous}. For both experiments, the data
are analysed with a maximum likelihood analysis followed by the extraction of
confidence levels in a unfied approach. It is not the task nor the purpose of 
this analysis to treat the apparent disagreement between the two experiments.
Rather, assuming no serious systematical error in either experiment or their 
interpretation with respect to \numubnueb\ oscillation, we apply a consistent 
statistical analysis to quantitatively establish their level of compatibility.
The work follows an earlier analysis \cite{NJP} based on intermediate data sets.

There are two separate questions investigated in this paper that relate
to the LSND and KARMEN data. The first question is: at what level are the
two data sets {\it compatible} with each other? This question addresses
whether or not the two experiments force one to draw opposite conclusions
from their respective data.

If one {\it assumes} that the data sets are compatible, the second
question becomes: what are the most likely regions for the oscillation
parameters? This method is somewhat analogous to treating the KARMEN
data as coming from a 'near detector' at 17 meters, and treating the LSND
data as coming from a 'far detector' at 30 meters. It ignores the question
of systematic differences between the experiments, such as whether or not,
for example, an unaccounted source of \nueb\ background neutrinos was
present in the LSND data and not present in the KARMEN data.

The paper is organized as follows: After a short overview of the experimental
setups of both experiments, section~\ref{sec:lhd-ana} describes the data analyses
leading to likelihood functions of the oscillation parameters. 
Section~\ref{sec:ind-conf} is devoted to the extraction of confidence regions for
the individual experiments. In section~\ref{sec:comp}, the
experimental results are combined statistically, extracting levels of 
compatibility as well as oscillation parameters compatible with both experiments.
We conclude with implications of this joint analysis, a comparison with other
oscillation searches and an outlook to upcoming experimental tests of the favored 
oscillation parameters presented in this analysis.
%
\subsection{\label{sec:intro_L}The LSND experiment}
%
The source of neutrinos for the 
LSND experiment was the interaction of the 798\,MeV proton beam at the
Los Alamos Neutron Science Center (LANSCE), in which a large number of pions, 
mostly $\pi^+$ are produced.  The $\pi^-$ are mainly absorbed and
only a small fraction decay to $\mu^-$, which in turn are largely
captured.  Thus, the resulting neutrino source is dominantly due to
\pipmup\ and \mupdecay\  decays, most of which decay at rest (DAR). With a very
small contamination of $\nueb/\numub \sim 8 \cdot 10^{-4}$, a measurement 
of the reaction \nuep\ provides a sensitive 
way to search for \numubnueb\ oscillations. Such events are
identified by detection of both the $e^+$ and the 2.2 MeV $\gamma$
from the reaction \pnd . In addition, the
$\nu_e$ flux from $\pi^+$ and $\mu^+$ decay-in-flight (DIF) is very
small, which allows a search for \numunue\ oscillations via the 
measurement of electrons above the Michel electron endpoint from 
the reaction \CC .

The LSND experiment took data over six calendar years (1993-1998).  
During this period the LANSCE accelerator operated for 17 months, delivering 
28\,896\,C of protons on the production target.  
The LSND detector \cite{bigpaper} consisted of an approximately
cylindrical tank 8.3\,m long by 5.7\,m in diameter.  The center of the
detector was located 30\,m from the beam stop neutrino source. The tank was filled 
with liquid scintillator consisting of mineral oil and 0.031 g/l of
b-PBD.  This low scintillator concentration allowed the detection of
both \cerenk\ light and scintillation light. Photomultiplier time and 
pulse-height signals were used to reconstruct the track with an average RMS
position resolution of $\sim 14$\,cm and an energy resolution of $\sim 7$\,\%
at the Michel endpoint of 52.8\,MeV.
%
\subsection{\label{sec:intro_K}The KARMEN experiment}
%
The KARMEN experiment used as neutrino source the pulsed spallation 
neutron source ISIS of the Rutherford Appleton Laboratory delivering 800\,MeV protons.
In contrast to the LANSCE source, the protons are extracted from the synchrotron 
as an intense but narrow double pulse, consisting of two parabolic
pulses with a width of 100\,ns separated by a peak-to-peak gap of 325\,ns.
The unique time structure of the ISIS proton pulses allowed a clear separation of
neutrino-induced events from any beam unrelated background.
The intrinsic contamination of the ISIS \numub --beam
is $\nueb / \numub = 6.4\cdot 10^{-4}$ \cite{Bur96}.

The KARMEN detector \cite{drexlin} was a rectangular high resolution liquid
scintillation calorimeter, located at a mean distance of 17.7\,m
from the ISIS target at an angle of 100 degrees relative to the
proton beam. The liquid scintillator was enclosed by a multilayer
active veto system and a 7000\,t steel shielding. The 65\,m$^3$ of
liquid scintillator consisted of a mixture of paraffin oil
($75\%$vol.), Pseudocumene ($25\%$vol.) and 2\,g/l
scintillating additive 1-phenyl-3-mesityl-2-pyrazoline.
The liquid scintillator volume was optically separated into 512
independent central modules. Gadolinium was implemented between 
the module walls for an efficient detection of thermal neutrons
\Gdng\ with on average 3 $\gamma$'s of energy $\sum E_\gamma=8$\,MeV.
The KARMEN detector as liquid scintillator calorimeter was
optimized for high energy resolution of $\sigma_E = 11.5\% /
\sqrt{E(\mbox{MeV})}$. Each event had energy,
time and position information, as well as the number of addressed
modules and their relative time differences.

The KARMEN\,2 experiment took data from February 1997 to March
2001. During this time, protons equivalent to a total charge of 9425 
Coulombs have been accumulated on the ISIS target. This corresponds 
to $2.71\cdot 10^{21}$ \numub\ from the ISIS beam stop target.
%
\section{\label{sec:lhd-ana}Data analysis and results as likelihood functions}
%
For both experiments, a \nueb\ signal from \numubnueb\ oscillations consists
of a spatially correlated delayed (\pos,n) sequence from \CCprot . Due to the 
different experimental techniques, the parameters to identify such sequences 
vary. Details of the LSND event reconstruction can be found in \cite{LSNDdet} and 
\cite{bigpaper}, the requirements for event sequences in KARMEN are described in 
\cite{K2paper}.

In LSND, a (\pos,n) sequence requires a prompt 'electron-like' event with
energy $E_e>20$\,MeV followed by a low energy $\gamma$--event.
The information about the delayed event is characterized via a likelihood
ratio $R_\gamma$: If within one millisecond after the initial event another event 
is recorded at distance $\Delta r\le 250$\,cm, the likelihood ratio $R_\gamma$ in
energy (PMT hits), time and distance of being a correlated \pn\ over an
accidental coincidence is calculated, otherwise $R_\gamma=0$.
Requiring a high likelihood ratio $R_\gamma$ selects (\pos,n) correlated events with
low uncorrelated background, the so-called ``gold-plated'' events. However,
to determine the oscillation parameters \sit\ and \Dm\ in a likelihood analysis, 
an event sample with a much looser cut in $R_\gamma$ is used, leading to higher 
efficiency for the oscillation channel, but naturally increasing the background.
For the LSND sample analysed here, all the cuts described in \cite{LSNDfinal} 
have been applied with two exceptions: The energy of the prompt event has been
restricted to $20<E_e<60$\,MeV instead of $20<E_e<200$\,MeV and we applied
a loose cut of $R_{\gamma}>10^{-5}$ instead of no $R_\gamma$-cut at all. 
The reason for these 
changes is twofold. First, the upper energy cut $E_e<60$\,MeV excludes any
significant influence of the \numunue\ channel on the much more sensitive 
\numubnueb\ channel. For any value of the oscillation parameter \Dm , the
expected contribution from \numunue\ remains below 2\,\%
of the \numubnueb\ signal. The second reason is a more technical one. 
Applying a unified frequentist approach to the likelihood function gained from
the reduced data implies simulation and analysis of a large number of event 
samples analog to the experiment. Therefore, one is interested not to have too
large samples containing mainly background events. With the cuts above, the 
original event sample is reduced by a factor of 5.5 to 1032 candidates, with 
an efficiency for \numubnueb\ at high \Dm\ reduced to 64.4\,\% of the original one
(only 3.2\,\% of the expected \numunue\ events remain).

Figure~\ref{fig_lsndevents} shows the event sample comprising 1032 beam-on events. 
Four variables are used to categorize the events: 
\begin{figure*}
\begin{center}
\includegraphics[width=11cm]{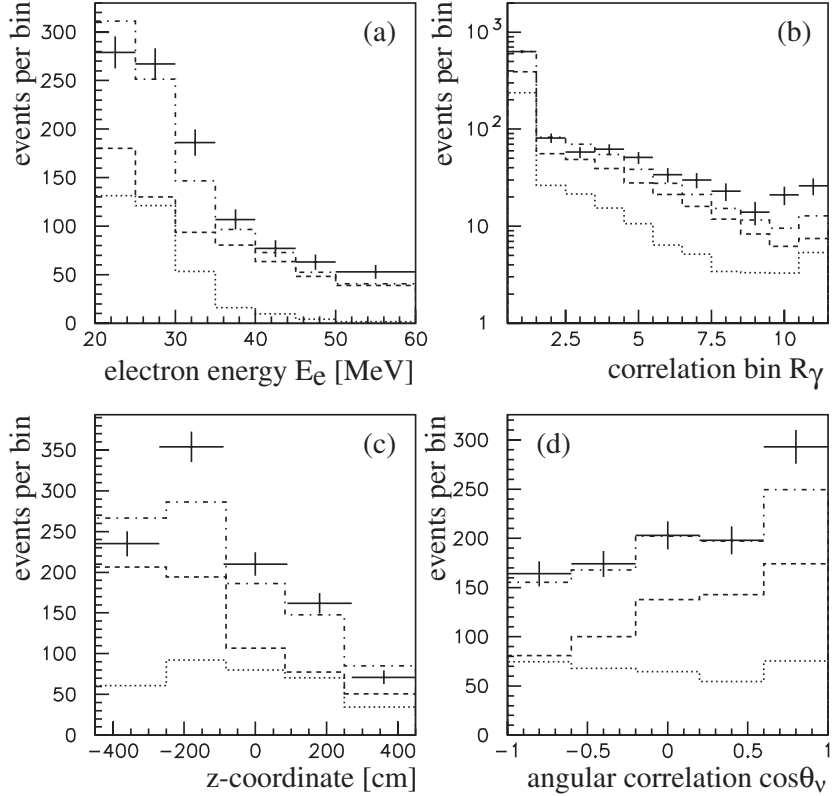} 
\caption{LSND data ensemble, together with the BUB (dashed), BRB (dotted) and total
background expectations BUB+BRB (dashed-dotted): (a) electron energy, (b) $R_\gamma$, (c)
spatial distribution along detector axis, and (d) directional angle $\cos \theta_\nu$.}
\label{fig_lsndevents}
\end{center}
\end{figure*}
The energy of the primary electron $E_e$, its spatial distribution along the detector
axis $z$ \footnote{The spatial coordinate $z$ along the detector axis can be easily
referred to the neutrino flight path $L$ by adding the distance source--center of 
detector tank.} and the angle $\cos \theta_\nu$ between the direction of the incident 
neutrino and the reconstructed electron path. The fourth variable is the likelihood 
ratio $R_{\gamma}$ for a (\pos ,n) coincidence. The binning of this distribution with 
the upper edge of bin number $N$ representing a value $R_{\gamma}=10^{N/3-1}$ has been
chosen for practical reasons. Note that the spectra in Figure~\ref{fig_lsndevents}
are projections of a 4-dim space of correlated parameters for each event.
Superimposed to the data are shown the beam-unrelated and beam-related background
contributions amounting to a total number of $N_{bgd}^{tot}=971.9$ events.
The expected background contributions are broken down in Table~\ref{tab:tb1}, 
for further details we refer to \cite{LSNDfinal}.
\begin{table*}
\begin{ruledtabular}
\caption{Contributions to the LSND data sample of different beam--related (BRB 3--6,8) 
and beam-unrelated (BUB) background processes as well as the expected oscillation
events for $\sit=1$, $\Dm=100$\,\eVc .}
\begin{tabular}{cccc}
Contribution & Signal or Background Source & Process &
Expected Number of Events \\ \hline
1 & \numubnueb & $\bar \nu_e p \rightarrow e+ n$ & 11350$\pm$115 \\
2 & BUB & & 635.0$\pm$26.2 \\
3 & DAR $\nu_e$ &
 $\nu_e\;^{12}\,C\rightarrow e^- N_{g.s.}$& 312.2$\pm$18.5 \\
&& $\nu_e\;^{12}\,C\rightarrow e^- N^*$& \\
&& $\nu_e\;^{13}\,C \rightarrow e^- N$& \\
&& $\nu e \rightarrow \nu e$& \\
4 & DIF $\nu_\mu$ & $\nu_\mu C\rightarrow \mu^- N^*$& 7.4 \\
&&  $\nu_\mu C\rightarrow \mu^- N_{g.s.}$& \\
5 & DIF $\overline\nu_\mu$ & $\overline\nu_\mu p \rightarrow \mu^+ n$&3.9 \\
&&  $\overline\nu_\mu C \rightarrow \mu^+ B^*$& \\
&&  $\overline\nu_\mu C\rightarrow \mu^+ B_{g.s.}$& \\
6 & DAR $\overline \nu_e$ ($\mu^-$ DAR)& $\overline\nu_e p \rightarrow e^+ n$&12.4 \\
7 & \numunue &$\nu_e C \rightarrow e^- N$&227$\pm$30 \\
8 & DIF $\pi^+\rightarrow\nu_e$ and $\mu^+\rightarrow\nu_e$ decay &
$\nu_e C \rightarrow e^- N$&1.0 \\
\end{tabular}
\label{tab:tb1}
\end{ruledtabular}
\end{table*}

Formally, each beam-on event $j$ of the $N_{beam-on}=1032$ candidates
is assigned a probability $p_j(\vec x)$ equal to a sum of probabilities
 $q_i(\vec x)$ from the backgrounds plus oscillations. The vector $\vec x$ 
describes herein the spectral parameters of each event $j$, 
$\vec x_j=(E_{ej},R_{\gamma j},\cos\theta_{\nu j},z_j)$.
It then remains to add the $q_i$ with expected
fractional contributions $r_i$ and take the product over all the
beam-on events. The likelihood is thus
\begin{equation}
\mathcal{L} =  (\prod_{j=1}^{N_{beam-on}} p_j),
\end{equation}
where
\begin{equation}
p_j(\vec x_j) = \sum_{i=1}^{N_{contr.}} q{_i}(\vec x_j) \cdot r_i.
\end{equation}
Additionally, two normalization requirements must hold:
\begin{equation}
\sum_{i=1}^{N_{contr.}} r_i = 1,
\label{equ:norm1}
\end{equation}
and
\begin{equation}
\int dE_e\,dR_\gamma\,d(\cos\theta_\nu)\,dz\,\, q_i(E_e,R_\gamma,\cos\theta_\nu,z) = 1
\label{equ:norm2}
\end{equation}
for each contribution, $i$.  Together, these requirements ensure that
every observed beam-on event has a probability of occurrence equal to 1.

\begin{figure*}
\begin{center}
\includegraphics{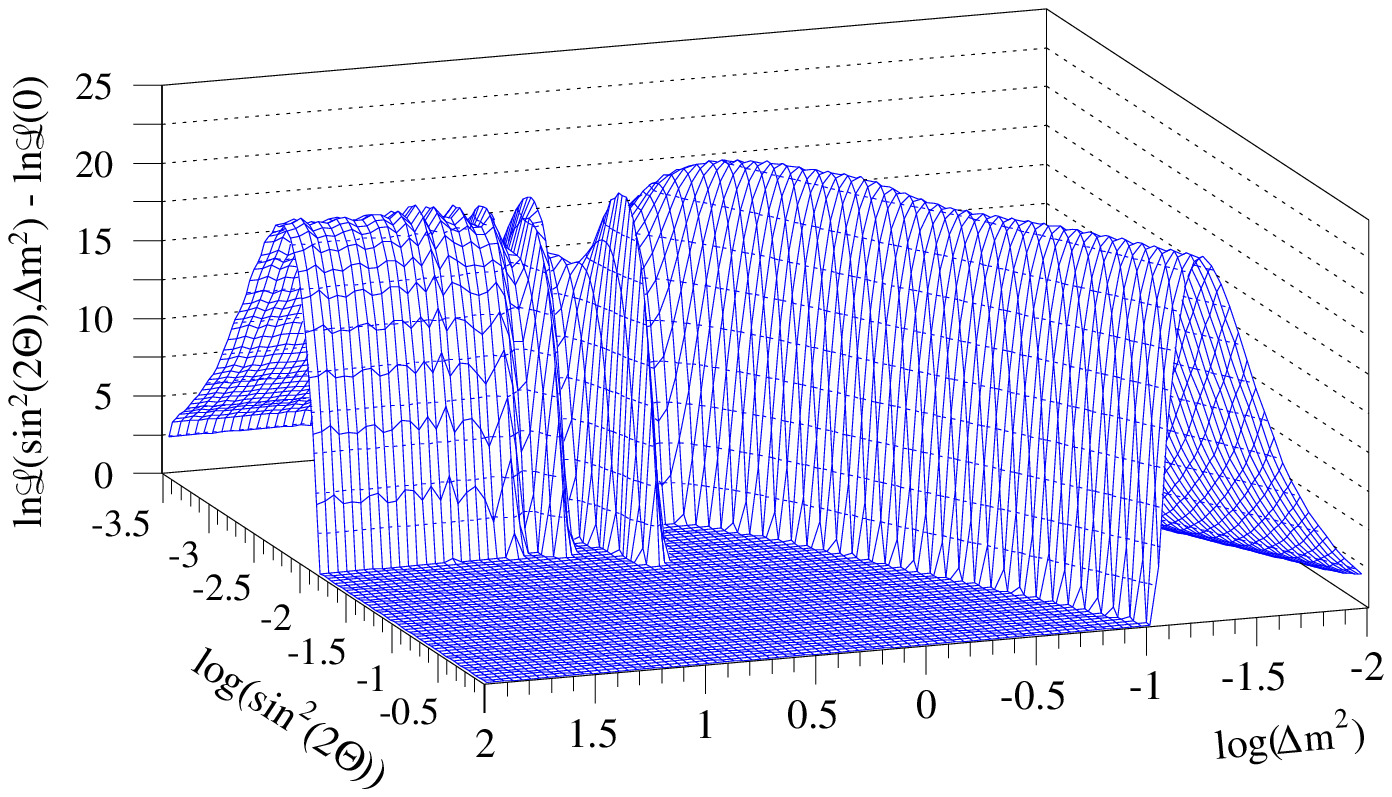} 
\caption{LSND logarithmic likelihood function $\ln \mathcal{L}(\sit,\Dm)$}
\label{fig_lsndlhd}
\end{center}
\end{figure*}

To allow for the fact that the backgrounds are known
to some limited accuracy only, the background is varied by calculating
the above likelihood at each point in the (\sit ,\Dm ) plane many times, 
varying over the expected $\sigma$ for each background.  
For each background configuration, the $\mathcal{L}$ is
weighted with a Gaussian factor for each background that is off its
central value.  The background variations are performed in a simplified manner,
i.e. the beam-unrelated background (BUB) varies independently and all the
beam-related backgrounds (BRBs) are locked together. Finally, the likelihood 
can be expressed as
\begin{equation}
\mathcal{L} = \int\mathcal{D}N_{bgd}
\exp(-(N_{bgd}-N_{bgd,exp})^2/2\sigma^2) \cdot \prod_{j=1}^{N_{beam-on}} p_j \ , 
\label{equ:bgvar}
\end{equation}
where the $\int\mathcal{D}N_{bgd}$ represents, schematically, the
background variation described above. Integrating over different background
contributions eventually reduces the free parameters of the likelihood
procedure from $N_{contr.}+1$ to the 2 free parameters of real
interest within this analysis, the oscillation parameters:
\begin{equation}
\mathcal{L}(r_1,\dots r_{N_{contr.}-1},\sit,\Dm) \rightarrow
\mathcal{L}(\sit,\Dm)
\end{equation}
Figure~\ref{fig_lsndlhd} shows the logarithm of the event--based likelihood 
function, $\lnL(\sit,\Dm)$, for the 1032 events investigated.
The maximum is reached at a parameter combination (\sit=0.85, \Dm=0.055\,\eVc)
corresponding to a total oscillation signal 
of $N_{osc}=N(\numubnueb)+N(\numunue)=66.4+0.3=66.7$. The difference to the 
no--oscillation hypothesis as expressed in logarithmic likelihood units is
\begin{equation}
\lnL(0.85,0.055\eVc)-\lnL(0,0)=23.5
\label{LSND_max}
\end{equation}
underlining the significance of the additional \nueb\ signal among the
event sample. Note that the extracted oscillation events $N_{osc}$ scale with
the efficiency of the applied cuts and are in good agreement with the best fit
result stated in \cite{LSNDfinal}.

\begin{figure*}
\begin{center}
\includegraphics[width=12cm]{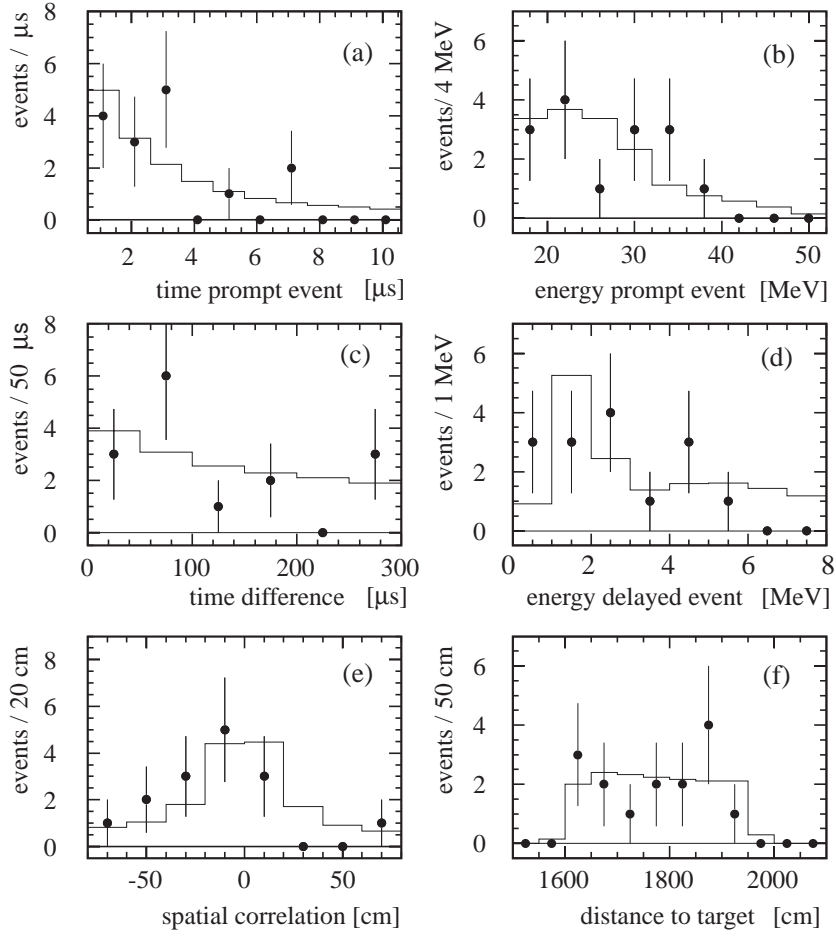} 
\caption{KARMEN\,2 event ensemble: (a) time of prompt events, (b)
energy of prompt events, (c) time difference between prompt and
delayed event, (d) energy of delayed events, (e) spatial
correlation and (f) distance to target of prompt event. The 15
oscillation candidates are in very good agreement with the
background expectation of 15.8 events (solid line).}
\label{fig_k2events}
\end{center}
\end{figure*}

KARMEN\,2 collected data from February 1997 through March 2001 corresponding
to 9425\,C accumulated proton charge on the ISIS target. A spatial 
coincidence between the initial \pos\ and the neutron capture of 1.3\,m$^3$ 
was required. Applying all cuts to the data \cite{K2paper}, 15 (\pos ,n) candidate 
sequences were finally reduced. Figure~\ref{fig_k2events} shows the remaining 
sequences in the appropriate energy, time and spatial windows.
The background components are also given with their distributions.
All components except the intrinsic \nueb\ contamination are measured
online in different time and energy windows (see Table~\ref{tab:bgsum}).
\begin{table*}
  \begin{ruledtabular}
  \caption{Expected KARMEN\,2 background components and \numubnueb\ signal}
  \label{tab:bgsum}
  \begin{tabular}{lcl}
   Process & Expectation & Determination\\  \hline
   Cosmic induced background & $3.9 \pm 0.2$ &  measured in diff. time window\\
   Charged current coincidences & $5.1 \pm 0.2$ &  measured in diff. energy, time windows\\
   \nue --induced random coincidences & $4.8 \pm 0.3$ &  measured in diff. time window\\
   \nueb\ source contamination & $2.0 \pm 0.2$ &  MC--simulation\\ \hline
   Total background $N_{bg}$ & $15.8 \pm 0.5$&   \\ \hline
   $N_{osc}(\sit=1,\Dm=100\,\eVc)$ & $2913\pm 267$ & \\
  \end{tabular}
  \end{ruledtabular}
\end{table*}
The extracted number of sequences is in excellent agreement with the background
expectation, consistent with no oscillation signal. To also include the detailed 
spectral information of each individual event, a maximum likelihood method is applied.

Analogously to the LSND likelihood function, we can define the combined likelihood
for the KARMEN sample as
\begin{equation}
\mathcal{L} =  (\prod_{j=1}^{N_{beam-on}} p_j),
\end{equation}
where
\begin{equation}
p_j(\vec x_j) = \sum_{i=1}^{N_{contr.}} q{_i}(\vec x_j) \cdot r_i
\label{equ:pjk2}
\end{equation}
with $N_{beam-on}=15$ and $N_{contr.}=5$, the oscillation signal as well 
as 4 background contributions.
For KARMEN\,2, the vector $\vec x$ has, of course, a different definition,
namely $\vec x=(E_{pr},T_{pr},E_{del},\Delta T,\Delta x)$ describing the energy
and time relative to beam-on-target of the prompt event, the energy of the 
delayed event as well as the time difference and spatial correlation of the prompt
and delayed event. Again, normalizations (\ref{equ:norm1}) and (\ref{equ:norm2})
with the appropriate event parameters have to hold. Differing from the likelihood
definition in the LSND analysis, for KARMEN with its very small background,
all background components are combined into one contribution, 
therefore reducing the above sum (\ref{equ:pjk2})
to only the oscillation signal and the total background, $N_{contr.}=2$.
Furthermore, the weight for the background variation, defined in (\ref{equ:bgvar}) as
a Gaussian factor, is taken as Gamma function, reflecting a continuous variation
in a Poisson statistics \cite{K2paper}.

\begin{figure*}
\begin{center}
\includegraphics{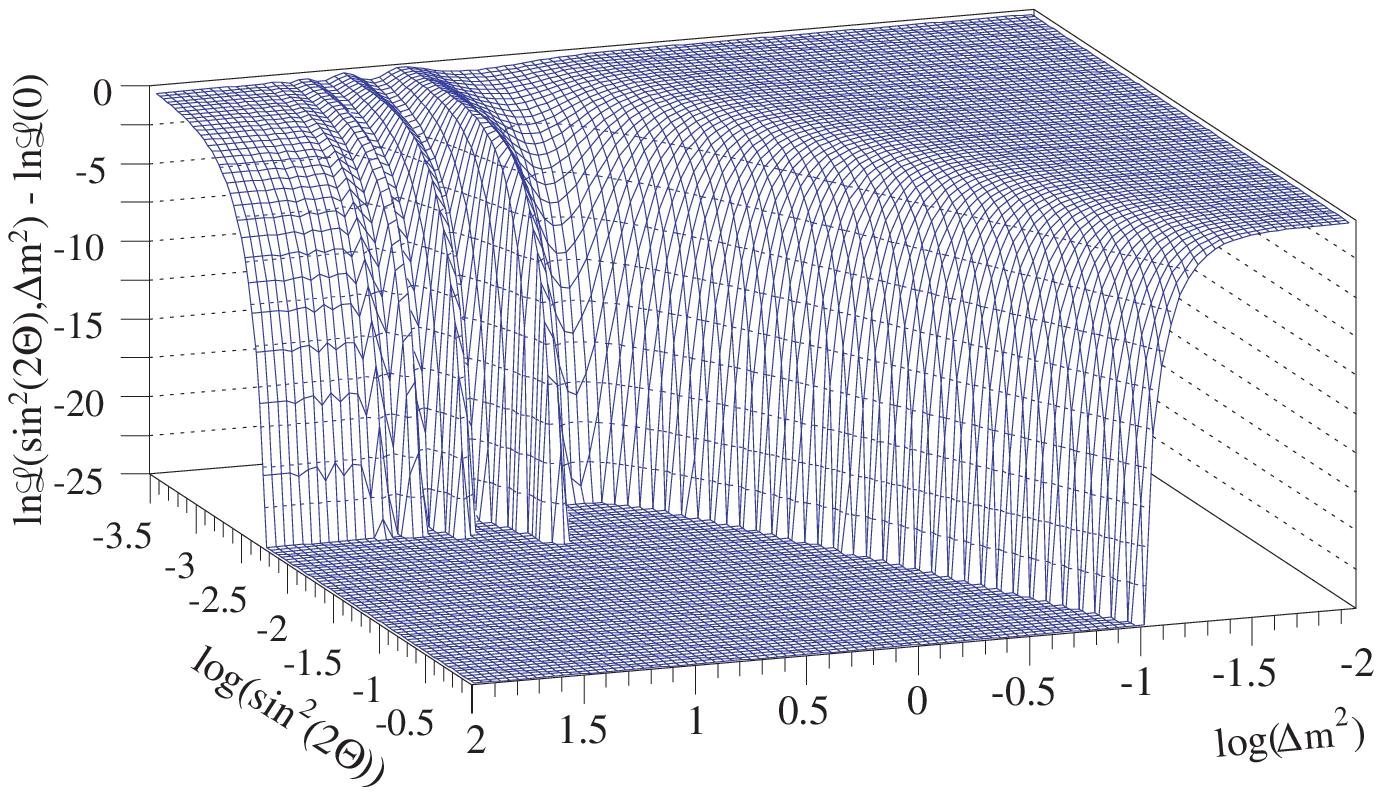} 
\caption{KARMEN\,2 logarithmic likelihood function $\ln \mathcal{L}(\sit,\Dm)$}
\label{fig_k2lhd}
\end{center}
\end{figure*}

Figure~\ref{fig_k2lhd} shows the logarithm of the likelihood as a function of the
oscillation parameters (\sit,\Dm). The maximum is reached near the physical boundary
 $\sit=0$, i.e. the no--oscillation case. The sharp drop in \lnL\ towards larger 
values of \sit\ demonstrates that there is no indication for an oscillation signal.

A typical approach to deduce confidence regions from likelihood functions
would consist of defining the contour of an area of given confidence
by cutting the likelihood function at the corresponding value below the maximum.
For a Gaussian shaped likelihood function of two uncorrelated free parameters, 
the corresponding values would be 
\begin{equation}
 \Delta \ln \mathcal{L}_{\rm Gauss}=-\ln(1-\delta/100)=2.3(4.6)
 \label{eq:Gauss}
\end{equation} 
for $\delta=90(99)\,\%$ confidence, respectively \cite{Lyons}. 
However, this method leads to correct coverage only in the Gaussian
case. As can be seen from Figures~\ref{fig_lsndlhd} and~\ref{fig_k2lhd} 
the likelihood functions have distinct features: 
An oscillatory behavior as a function of the free parameters, 
the maximum at the physical boundary (for KARMEN), or numerous side maxima with
comparable likelihood values (for LSND). In addition, the parameter space 
in (\sit,\Dm) is, in principle, not limited, its metric in terms of prior
probability density not unequivocally defined.
%
\section{\label{sec:ind-conf}Individual confidence regions}
%
The method we apply in the following is based on a unified frequentist approach
suggested by \cite{feldcous} which eliminates the bias that occurs when one decides,
after analysing the data, between using a confidence interval (having a positive 
signal) or an upper confidence limit (having a result compatible with the background).
When first presented, it was argued that the suggested ordering principle near the 
physical boundary should be modified \cite{Giunti},\cite{Byron}, or that a 
Bayesian approach to the extracted likelihood function would be more 
appropriate \cite{Dagos}. However, many experimental results have been analysed 
since using the unified approach \cite{CWCL}, also being described as a standard 
procedure by \cite{Pdg2000}.

The basic idea of the unified frequentist approach is to create a large number of event 
samples in perfect analogy to an experiment. These samples are created by Monte Carlo
using the full event information for the likelihood procedure.
For an oscillation hypothesis $H$ with given parameters (\sit,\Dm)$_H$,
a large sample of Monte Carlo simulations of so-called toy experiments is 
created. These simulations are based on the detailed knowledge of all 
experimental resolution functions and the spectral information on the 
individual background contributions. In addition, they comprise the
expected experimental signal for the given oscillation hypothesis \footnote{For 
each individual sample, the number of background events as well as the number 
of signal events vary according to the error of the expectation values.}.
For each likelihood function of the analysed samples, the estimator
 $\Delta \ln \mathcal{L}= \ln \mathcal{L}_{max} - \ln \mathcal{L}_H$
is calculated by comparing each sample's likelihood maximum with the value 
for the given input parameters (\sit,\Dm)$_H$. 
The hypothesis $H$ is then accepted at a confidence $\delta$,
if the estimator $\Delta \ln \mathcal{L}$ of the experimental likelihood
function is contained within the smallest $\delta$\,\% 
of the simulated estimator distribution. Since, in principle, the
estimator distribution itself is a function of the oscillation parameters,
a complete statistical analysis consists of a scan of the entire parameter space
(\sit,\Dm) to extract the according region of confidence.
%
\subsection{\label{sec:ind_K}KARMEN}
%
To deduce confidence regions from the KARMEN\,2 likelihood function,
the parameter space (\sit,\Dm) has been scanned extracting the estimator 
distribution for 8000 different oscillation hypotheses, including the 
no--oscillation scenario. From these distributions, each containing the 
information of about 4000 simulated and analysed samples, regions of different 
confidence levels can be calculated.
\begin{figure}
\begin{center}
\includegraphics{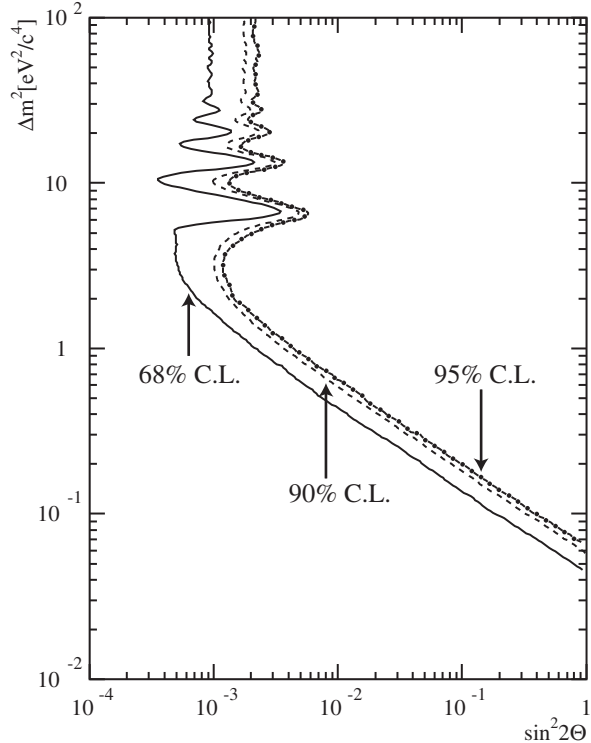} 
\caption{KARMEN\,2 regions of various confidence. Parameter combinations
 to the right contain the complementary confidence.}
\label{fig_k2osciplot}
\end{center}
\end{figure}
Figure~\ref{fig_k2osciplot} shows the regions of the oscillation parameters
(\sit,\Dm) for confidence levels of 68\,\%, 90\,\% and 95\,\%.
Parameter combinations to the right contain the complementary confidence.
The contours of the confidence areas are often referred to as exclusion curves 
of a given confidence. For values $\Dm \ge 100$\,\eVc , the contours cut
at oscillation amplitudes of $\sit=0.9\cdot 10^{-3}$ (68\,\% C.L.),
 $\sit=1.7\cdot 10^{-3}$ (\NCL) and $\sit=2.2\cdot 10^{-3}$ (\NnCL).

The KARMEN\,2 sensitivity, defined as expectation value for the upper limit of
a given confidence interval under the assumption of no oscillations, is 
again obtained by simulations: 
 $\langle \sit \rangle =0.6(1.6,2.2)\cdot 10^{-3}$ for 68\,\% C.L. 
(\NCL, \NnCL), respectively. The error on the contour lines
is given by the statistical error of the estimator's cut value
 $\Delta \ln \mathcal{L}(\delta)$ and therefore depends on the amount of 
simulated samples per hypothesis and the degree $\delta$ of confidence under
consideration. For example, determining $\Delta \ln \mathcal{L}(90\,\%)$
from a distribution of 1000 simulated samples relies on the upper 10\,\% 
tail or the highest 100 $\Delta \ln \mathcal{L}$ values. We derive the error on
 $\Delta \ln \mathcal{L}$ from the spread of the cut value taking the upper 
 $100\pm\sqrt{100}$ samples. For KARMEN\,2, the typical relative error on
 $\Delta \ln \mathcal{L}(90\,\%)$ 
is in the range of $2-3$\,\%.
%
\subsection{\label{sec:ind_L}LSND}
%
For the LSND likelihood function, the method to deduce the confidence regions is
identical to the one described in section~\ref{sec:ind_K} for KARMEN\,2. 
Technically, however, there are differences due to the large amount of computing time.
This arises from the fact that the event samples are much larger (about 1000 events,
depending on the oscillation parameters, instead of about 15) and the integration over 
different background contributions according to equation~(\ref{equ:bgvar}) is performed 
in as much as 255 discrete steps of different backgrounds BUB and BRB.
\begin{table*}
  \begin{ruledtabular}
  \caption{LSND estimator values $\Delta \ln \mathcal{L}$ for 3 different confidence
  levels $\delta$ and varying input oscillation signal strength \sit\ but fixed 
  parameter $\Dm=1$\,\eVc .}
  \label{tab:lsndgrid}
  \begin{tabular}{ccccc}
   \sit & $N_{osc}$ & \multicolumn{3}{c}{$\Delta \ln \mathcal{L}$} \\  
   & DAR+DIF & 68\,\% & 90\,\% & 95\,\% \\ \hline
   0.0                & 0.0       & $2.25\pm 0.2$ & $4.25^{+0.3}_{-0.5}$ & 
   $5.45^{+1.2}_{-1.0}$ \\
   $2.5\cdot 10^{-3}$ &  33.5+0.2 & $2.55\pm 0.1$ & $4.05^{+0.7}_{-0.4}$ & 
   $5.15^{+1.8}_{-0.7}$ \\
   $5\cdot 10^{-3}$   &  67.1+0.4 & $2.65\pm 0.1$ & $4.25^{+0.5}_{-0.4}$ & 
   $5.35^{+1.1}_{-0.8}$ \\
   $1\cdot 10^{-2}$   & 134.2+0.8 & $1.95\pm 0.1$ & $3.55^{+0.7}_{-0.4}$ & 
   $4.55^{+1.9}_{-0.9}$ \\
  \end{tabular}
  \end{ruledtabular}
\end{table*}
Therefore, we restricted the hypotheses tested from a fine grid for KARMEN\,2 to 
some grid points representing the typical variation of expected oscillation events.
For a fixed oscillation parameter $\Dm=1$\,\eVc, 4 oscillation scenarios were
simulated with 1000 individual samples each (see Table~\ref{tab:lsndgrid}).

\begin{figure}
\begin{center}
\includegraphics{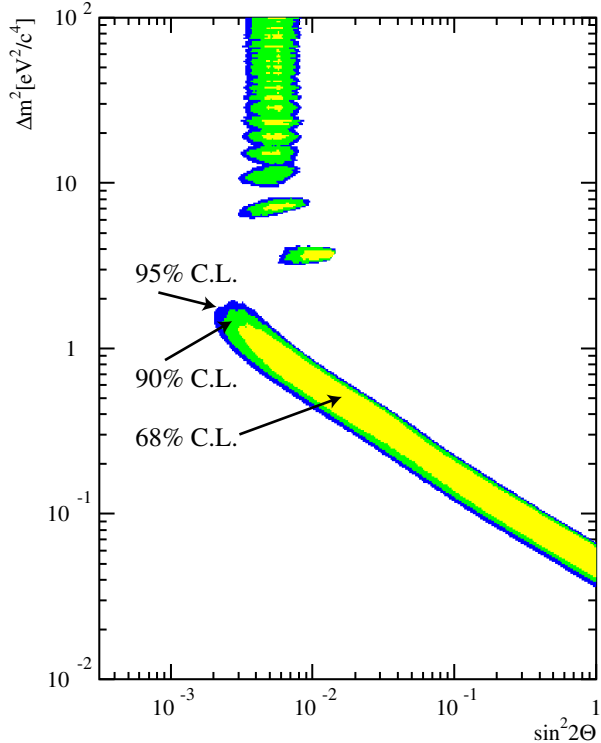} 
\caption{LSND regions of various confidence. Parameter combinations
 outside the contours contain the complementary confidence.}
\label{fig_lsndosciplot}
\end{center}
\end{figure}

As verified by earlier calculations \cite{NJP} and the KARMEN\,2 analysis, the
main variation of $\Delta \ln \mathcal{L}(\delta)$ for a given confidence $\delta$
in the parameter space (\sit,\Dm) follows the variation of the signal strength, 
i.e. the number of oscillation events $N_{osc}$ among the event sample.
We therefore extrapolated the values of $\Delta \ln \mathcal{L}$ in 
Table~\ref{tab:lsndgrid} linearly with $N_{osc}$ and adopted, at each grid point
(\sit,\Dm) with its expected number of oscillation events $N_{osc}$, the estimator
value $\Delta \ln \mathcal{L}(\sit,\Dm)$. Having constructed the estimator cut
distribution this way, the last step is analogous to that used for KARMEN\,2, 
i.e. to cut the 
LSND likelihood function with the given values $\Delta \ln \mathcal{L}(\sit,\Dm)$
below the maximum and draw the contour lines. Figure~\ref{fig_lsndosciplot} shows
the contour lines of areas containing 68\,\%, 90\,\% and 95\,\% confidence.
Although parameter combinations with $\Dm > 10$\,\eVc\ are part of the \NCL\
area, they are scarcely included in the more stringent interval of 68\,\% 
confidence, demonstrating the lower likelihood values for such oscillation 
scenarios.

It is useful to compare the extracted confidence regions with the favored
regions given in \cite{LSNDfinal} where the likelihood function had been calculated 
based on a larger event sample (5697 events with $20<E_e<200$\,MeV and 
no $R_\gamma$--cut). In addition, a simplified extraction of confidence regions
according to equation~(\ref{eq:Gauss}) had been applied
\footnote{In fact, due in part to the larger statistical sample, in \cite{LSNDfinal} 
the results of such a simplified procedure were shown to be similar to the results 
obtained from an approximation to a full unified frequentist approach.}.
\begin{figure}
\begin{center}
\includegraphics{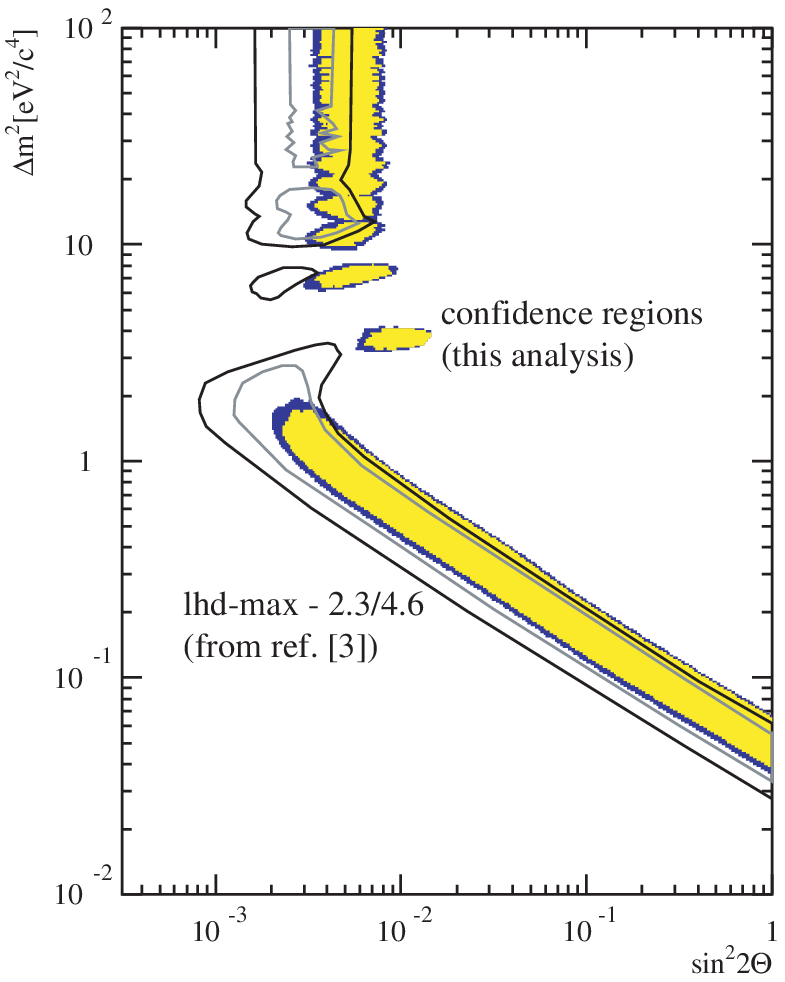} 
\caption{LSND regions of 90\,\% and 95\,\% 
 confidence in comparison with the favored region
 analysing the likelihood function of a larger event sample and applying a constant cut 
 $\Delta \ln \mathcal{L}=2.3(4.6)$, from \protect\cite{LSNDfinal}}
\label{fig_lsndnewandold}
\end{center}
\end{figure}
Figure~\ref{fig_lsndnewandold} shows, that for lower values of \Dm , there is only 
a slight shift to larger oscillation amplitudes of the new analysis. A large overlap 
of the \NCL\ area with the region according to $\Delta \ln \mathcal{L}=2.3$
demonstrates the good statistical agreement of the analyses of both event samples.
For higher values of \Dm , the differences are more pronounced. Compared to the older
analysis, the confidence region is not only shifted considerably to larger amplitudes,
but the likelihood values themselves are larger relative to the likelihood maximum
which results in a larger area of \NCL . These effects can be understood by the upper
energy cut $E_e<60$\,MeV applied in the actual analysis. Restricting the analysis
mainly to the oscillation channel \numubnueb\ favors also large values of \Dm\
whereas a signal from \numunue , derived in \cite{LSNDfinal} to be much smaller, 
but detectable mainly at higher energies, decreases the likelihood for such solutions 
as well as the absolute strength or oscillation amplitude.

Finally, we note the importance to derive values of $\Delta \ln \mathcal{L}$ leading
to correct coverage of confidence regions. These values differ considerably from the
simplified Gaussian approach of a constant of 2.3 units. In an earlier analysis of 
the LSND data \cite{NJP}, the typical estimator cut values were 
 $\Delta \ln \mathcal{L}(90\,\%)\approx 2.5-3.5$ on the basis of about 3000 events.
Now, with even smaller event samples, the values of $\Delta \ln \mathcal{L}$ further
increase to $\approx 3.5-4.5$. This dependance of $\Delta \ln \mathcal{L}$ is also 
underlined by the typical values for the much smaller KARMEN\,2 sample size, where
 $\Delta \ln \mathcal{L}(90\,\%)$ ranges from 3 to 5.
%
\section{\label{sec:comp}Compatibility Analysis}
%
Having analysed both experiments' likelihood functions with the same consistent
method, we can now use this method and its results to deduce quantitative statements
on the question of statistical compatibility of both experimental results and,
in the case of such compatibility, on the common parameter combinations (\sit,\Dm).
%
\subsection{\label{sec:complevel}Level of compatibility}
%
One of the most misleading but nevertheless very frequently used 
interpretation of the LSND and KARMEN results is to take the LSND region left 
over from the KARMEN exclusion curve as area of (\sit,\Dm) favored by both experiments. 
Such an interpretation, though appealingly straight forward, completely ignores 
the information of both likelihood functions and reduces them to two discrete 
levels of a specific confidence $\delta$. To be able to correctly combine the 
two experimental results and extract the combined confidence regions,
we have to use the original estimator distributions for KARMEN and for LSND.
In a first step, the level of statistical compatibility of the two experimental 
outcomes has to be quantified. In the second step, the parameters favored by
both experiments will be deduced.

In the following analysis, an {\it experiment} or {\it event}, in logical 
terms of frequency or probability of occurence, means repeating both, 
the KARMEN {\it and} the LSND experiment. Compatibility is then achieved if 
there are at least some oscillation parameters being the element of both the 
repeated KARMEN-like and LSND-like experiment's confidence region.

We make the well justified assumption that the two experiments LSND and KARMEN are 
independent. Then, a two dimensional estimator distribution can be constructed
for each hypothesis $(\sit,\Dm)_H$ from the individual distributions $\DlnL_K$ of 
KARMEN and $\DlnL_L$ of LSND by an inverse projection. 
For each hypothesis, the pair of experimental values
 $(\DlnL_{KARMEN2},\DlnL_{LSND})$ is checked to be part of the 'inner' $\delta$\,\% 
of the simulated 2--dimensional estimator distribution. There are different methods 
of ordering these 2--dimensional distributions all leading to very similar 
results \cite{NJP}. In the following, we require each experimental
value $\DlnL_{KARMEN2}$ and $\DlnL_{LSND}$ to be contained in the 
corresponding 1--dimensional interval of confidence. This selection criterion is 
equivalent to selecting a rectangle of frequency of the 2--dim estimator 
distribution. The resulting confidence is $\delta=\delta_{KARMEN}\cdot\delta_{LSND}$.
Since we do not weight the experiments, we define $\delta_{KARMEN}=\delta_{LSND}$,
so that for a combined confidence of e.g. 81\,\%, 
we demand $\DlnL_{KARMEN2}<\DlnL_K(90\,\%)$ and $\DlnL_{LSND}<\DlnL_L(90\,\%)$.

At a level of $\delta_{KARMEN}=\delta_{LSND}=60\,\%$, 
no parameter combination (\sit,\Dm) fulfills the requirement 
 $(\DlnL_{KARMEN2}<\DlnL_K) \wedge (\DlnL_{LSND}<\DlnL_L)$ any more. 
This corresponds to the fact that, at the level of combined confidence
\begin{equation}
 \delta_{inc}=0.6^2=36\,\% \ , 
\label{eq:incomp}
\end{equation}
the two experiments are completely incompatible.
Coming back to our definition of probability of occurence, in 64\,\%
of repetitions of double experiments, a KARMEN\,2 plus an LSND experiment with their
typical parameters and statistics, the outcome would consist of statistically
compatible results with a specified set of oscillation parameters (\sit, \Dm).
%
\subsection{\label{sec:favored}Common oscillation parameters}
%
In the following, we analyse the preferred regions of oscillation parameters in the case of
compatibility defined in the section before. If two experiments are independent 
producing two likelihood functions as result of the event analysis, the recipe for a 
combined analysis is straight forward. The likelihoods can be multiplied, i.e. in our 
case the logarithms of the likelihood functions are added. The absolute values of the
likelihood functions are somewhat arbitrary. A recommended presentation of \lnL\ 
is to normalize the individual functions $\lnL_{KARMEN}$ and $\lnL_{LSND}$ to a 
point in (\sit,\Dm) where they are equally sensitive to a potential signal \cite{dagos2}. 
In our case of the oscillation search this corresponds to values of $\sit=0$.
Therefore we define the combined likelihood function
\begin{widetext}
\begin{eqnarray} \nonumber
 \lnL(\sit,\Dm) &=& \left\{ \lnL_{KARMEN2}(\sit,\Dm)
                 - \lnL_{KARMEN2}(\sit=0) \right\}  \\ 
	        &+& \left\{ \lnL_{LSND}(\sit,\Dm)
                 - \lnL_{LSND}(\sit=0) \right\} \ .
\label{kl_lhd}
\end{eqnarray}
\end{widetext}
\begin{figure*}[hbt]
\begin{center}
\includegraphics{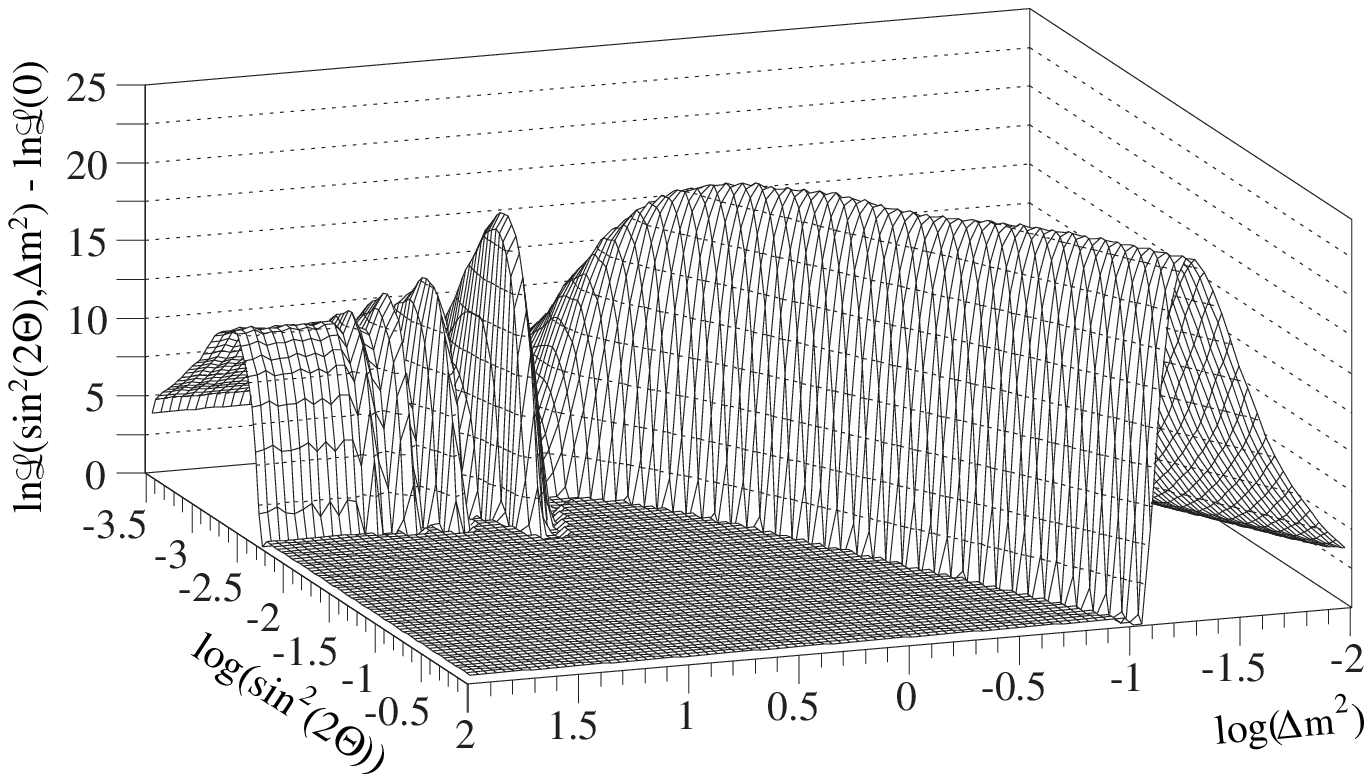} 
\caption{Combined KARMEN\,2 and LSND logarithmic likelihood function $\lnL(\sit,\Dm)$}
\label{kl_lhdfunction}
\end{center}
\end{figure*}
Figure~\ref{kl_lhdfunction} shows the combined function $\lnL(\sit,\Dm)$
with a maximum of \mbox{\lnL(\sit$=$1,\Dm$=$0.05)$=$21.5} on a long flat 'ridge'
of low \Dm\ values. The positive signal of LSND dominates the combined
likelihood. However, the maximum in comparison to the no--oscillation value
is reduced (see Equ.~\ref{LSND_max}) as well as the shape at higher values of \Dm\ 
is significantly modified due to the KARMEN\,2 likelihood function.

To determine the confidence regions in (\sit,\Dm), again we apply the unified approach,
i.e. we create the estimator distribution \DlnL$_{K+L}$ for various hypotheses 
(\sit,\Dm)$_H$ and compare the experiment's value \DlnL$_{KARMEN2+LSND}$ with 
the cut value \DlnL$_{K+L}(\delta)$ for a given confidence. To get the estimator,
a simulated event sample now  consists of a KARMEN\,2-like and an LSND-like 
MC sample each analysed with the appropriate likelihood definitions. Then both 
likelihood functions are added following equation~(\ref{kl_lhd}) and the estimator
 $\DlnL_{K+L}=\lnL_{K+L}(\sit,\Dm)_{max}-\lnL_{K+L}(\sit,\Dm)_H$ is calculated.
\begin{figure}
\begin{center}
\includegraphics{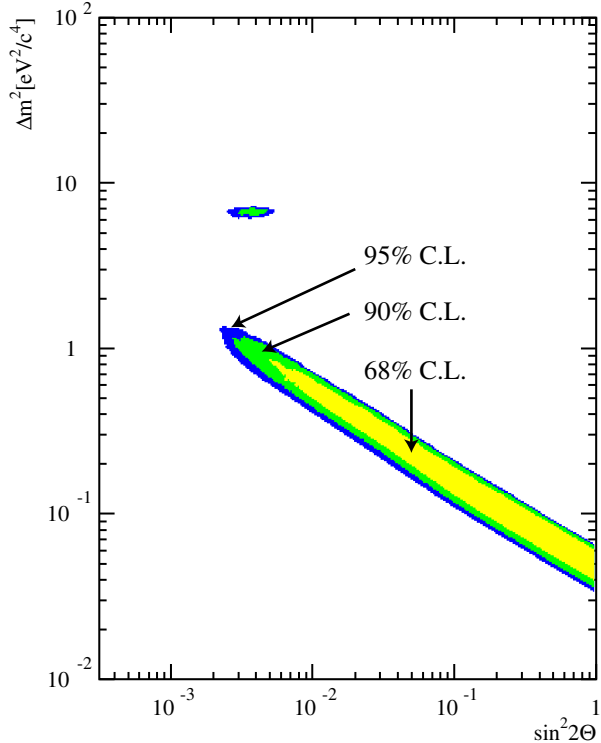} 
\caption{Regions of various confidence for the combined analysis assuming statistical
	 compatibility of KARMEN\,2 and LSND.}
\label{fig_combosciplot}
\end{center}
\end{figure}
The typical cut values for a confidence of $\delta=90$\,\% are slightly
higher than the individual LSND ones, ranging from 3.5 to 5.5.
Figure~\ref{fig_combosciplot} shows the confidence regions of the oscillation parameters
for the combined likelihood analysis. The total confidence of a parameter region
is hereby reduced by the fraction of incompatibility of the two experiments, so
\begin{equation}
 \delta_{tot} = \delta \cdot (1 - \delta_{inc})
 \label{eq:confidence}
\end{equation} 
which results in e.g. $\delta_{tot} = 0.9\cdot 0.64= 58$\,\% 
total confidence for the parameter combinations within the area denoted as 
\NCL\ area in Figure~\ref{fig_combosciplot}.

The combination of both experiments' results with this consistent frequentist approach
demonstrates that \numubnueb\ solutions with $\Dm>10$\,\eVc\ are excluded.
There remain essentially two solutions in the parameter space of \numubnueb\
oscillations, one at $\Dm \approx 7$\,\eVc\ and the area with $\Dm<1$\,\eVc .
The latter one, though graphically rather large in Figure~\ref{fig_combosciplot}
corresponds to just one solution in the 'phase space' of oscillations:
Any \nueb\ signal from oscillation has rather low energy, the evolution of the
oscillation probability (equ.\,\ref{eq:Osci}) being just at the beginning of the 
oscillation length. With an oscillation parameter $\Dm=0.5$\,\eVc, and a 
typical lower energy of $E_{\nu}=20$\,MeV from \mup\ decay at rest, the
first maximum of oscillations of \numub\ into \nueb\ would be at a distance
\begin{equation}
 d_{1.max} = \frac{\pi}{2}\cdot \frac{E_{\nu}}{1.27\cdot \Dm} \approx 50\,{\rm m}
 \label{eq:L_osc}
\end{equation}
from the neutrino source.
For such oscillation parameters, the negative result of KARMEN\,2 compared to
the excess of LSND mainly reflects the different detector distances 
 $d_{KARMEN}\approx 17$\,m versus $d_{LSND}\approx 30$\,m from the source.
This is one way to reconcile the results of the two experiments.

To summarize, at a joint confidence level 
of $\delta_{tot}=0.36+0.64\cdot0.68\approx 80$\,\%, 
the two experiments are either incompatible or lead to just one single oscillation
solution at $\Dm<1$\,\eVc . The remaining 20\,\% 
confidence separate into an enlargement of the parameter region of low \Dm\ values 
and into a second oscillation solution at $\Dm \approx 7$\,\eVc .
%
\section{\label{sec:conclu}Conclusion}
%
We have analysed both the KARMEN\,2 and the LSND final data with a maximum likelihood
method using a similar event--based likelihood function. The likelihood functions
as functions of the free parameters \sit\ and \Dm\ demonstrate the different outcomes
of the experiments, namely a clear, statistically significant excess of \nueb\ in LSND
versus no indication of an oscillation signal in KARMEN\,2. To deduce regions of 
correct coverage, we applied a unified frequentist approach to both likelihood 
analyses individually. The results underline the feasibility of as well as the 
necessity for such an approach.

A quantitative joint statistical analysis has been performed leading to a level
of 36\,\% incompatibility of the experimental outcomes, corresponding to
individual confidence levels of 60\,\%.
For the cases of statistical compatibility, the common parameter regions have been
identified on the basis of the unified frequentist approach applied to the
combined likelihood function of KARMEN\,2 and LSND. The derived confidence regions
in (\sit,\Dm) clearly differ from an often applied but incorrect graphical
overlap of the confidence regions of the individual experiments. 
There are two oscillation scenarios with either $\Dm\approx 7$\,\eVc\ or $\Dm<1$\,\eVc\
compatible with both experiments.

We performed a joint statistical analysis incorporating some of the systematic
uncertainties of the experiments, such as neutrino flux uncertainty, accuracy of
known cross sections and resolution functions of both experiments.
\begin{figure}
\begin{center}
\includegraphics{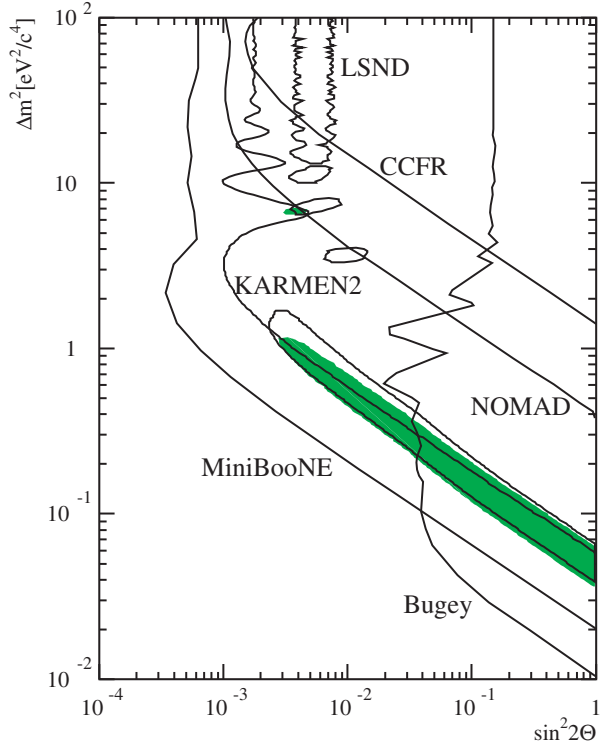} 
\caption{Parameter regions deduced in this work (grey area) compared with existing limits of
 experiments (Bugey \nuebx\ \protect\cite{Bugey}, CCFR \numunue\ \protect\cite{CCFR}
 and NOMAD \numunue\ \protect\cite{NOMAD}) and the envisaged sensitivity of the
 MiniBooNE experiment (with final single horn design \protect\cite{BooNE}).}
\label{fig_allosciplot}
\end{center}
\end{figure}
Further --unknown-- systematic uncertainties might become evident only by performing 
new experiments to confirm or discard the results described here. 
Figure~\ref{fig_allosciplot} shows 
the intended sensitivity of a new experiment, MiniBooNE at Fermilab \cite{BooNE},
which is under construction and will independently crosscheck the LSND evidence. 

In addition, we did not incorporate the results of other experiments on the 
oscillation parameters investigated. As shown in Figure~\ref{fig_allosciplot},
the deduced favored region in (\sit,\Dm) is partly overlaped by \NCL\ exclusion curves 
of other experiments. In particular, the recent NOMAD \numunue\ result \cite{NOMAD}
clearly reduces the overall likelihood for the $\Dm \approx 7$\,\eVc\ solution.
A complete analysis should include these results on the basis of the same statistical 
method, i.e. a consistent frequentist analysis. This implies, however, the
detailed knowledge of experimental data and resolution functions of these 
experiments not accessible to us. Furthermore, the exclusion curve from the 
Bugey experiment is based on the disappearance search \nuebx . Combining this 
experiment correctly with the appearance results of \numubnueb\ or \numunue\ 
in terms of mixing angles would therefore also require a full three-- or 
four--dimensional mixing scheme with theoretical models including sterile 
neutrinos \cite{bilenky,Hirsch,Ioann,Ma,Lorenz,strumia} or CPT violation \cite{barenb} 
and would become model-dependent.
%
\section{\label{sec:ackno}Acknowledgements}
%
We would like to thank the members of both collaborations for the encouragement and
help to make this final joint analysis possible. 
In particular, we acknowledge the support of and many discussions with W.C. Louis 
concerning the LSND as well as G. Drexlin concerning the KARMEN experiment.  
Without the openness and trust 
to analyse both data sets in detail, such an analysis wouldn't have been feasible.
K.E. also wants to thank the Alexander von Humboldt foundation for financial support
during visits to the Los Alamos National Laboratory. Clarifying discussions with 
Kai Zuber on the principles of combining experiments with contradictory central 
statements were very helpful.
%
%
%
\clearpage

%
%
\end{document}